\documentclass[sigconf,pbalance]{acmart}

\AtBeginDocument{%
  }

\copyrightyear{2024} 
\acmYear{2024} 
\setcopyright{rightsretained} 
\acmConference[ESEM '24]{Proceedings of the 18th ACM / IEEE International Symposium on Empirical Software Engineering and Measurement}{October 24--25, 2024}{Barcelona, Spain}
\acmBooktitle{Proceedings of the 18th ACM / IEEE International Symposium on Empirical Software Engineering and Measurement (ESEM '24), October 24--25, 2024, Barcelona, Spain}\acmDOI{10.1145/3674805.3695407}
\acmISBN{979-8-4007-1047-6/24/10}

\usepackage{cleveref}
\usepackage{siunitx}
\usepackage{booktabs}
\usepackage{csquotes}
\usepackage{todonotes}
\setuptodonotes{inline}
\usepackage{caption}
\usepackage{subcaption}
\usepackage{color}

\newcommand{\framed}[2]{%
  \vspace{0.5\baselineskip plus 1pt}
  \noindent\fbox{%
  \begin{minipage}{\linewidth-17pt}%
  \textbf{#1: }%
  #2%
  \end{minipage}}%
  \vspace{0.5\baselineskip plus 2pt}
}

\graphicspath{{./figures/}}

\begin{document}

\title[Do Test and Environmental Complexity Increase Flakiness?]{\texorpdfstring{Do Test and Environmental Complexity Increase Flakiness?\\An Empirical Study of SAP HANA}{Do Test and Environmental Complexity Increase Flakiness? An Empirical Study of SAP HANA}}

\author{Alexander Berndt}
\email{alexander.berndt@sap.com}
\orcid{0009-0009-5248-6405}
\affiliation{%
  \institution{SAP}
  \city{Walldorf}
  \country{Germany}
}
\author{Thomas Bach}
\email{thomas.bach03@sap.com}
\orcid{0000-0002-9993-2814}
\affiliation{%
  \institution{SAP}
  \city{Walldorf}
  \country{Germany}
}
\author{Sebastian Baltes}
\email{sebastian.baltes@uni-bayreuth.de}
\orcid{0000-0002-2442-7522}
\affiliation{%
  \institution{University of Bayreuth}
  \city{Bayreuth}
  \country{Germany}
}


\begin{abstract}
\emph{Background}: Test flakiness is a major problem in the software industry. Flaky tests fail seemingly at random without changes to the code and thus impede continuous integration (CI). Some researchers argue that all tests can be considered flaky and that tests only differ in their frequency of flaky failures. This position implies that the definition of test flakiness includes failures caused by interruptions in the testing environment. 

\emph{Aims}: With the goal of developing mitigation strategies to reduce the negative impact of test flakiness, we study characteristics of tests and the test environment that potentially impact test flakiness.

\emph{Method}: We construct two datasets based on SAP HANA's test results over a 12-week period: one based on production data of the SAP HANA CI pipeline, the other based on targeted test executions from a dedicated flakiness experiment. We conduct correlation analysis for test and test environment characteristics with respect to their influence on the frequency of flaky test failures. 

\emph{Results}: In our study, the average test execution time had the strongest positive correlation with the test flakiness rate ($r = 0.79$), which confirms previous studies. Potential reasons for higher flakiness include the larger test scope of long-running tests or test executions on a slower test infrastructure.
We found that distributed tests had a lower flakiness rate than non-distributed tests. Interestingly, the load on the testing infrastructure was not correlated with test flakiness. The relationship between test flakiness and required resources for test execution (i.e., memory and CPU) is inconclusive. % In one dataset, we found there exists a significant positive correlation. On the other dataset, we do not obtain a significant correlation. 

\emph{Conclusions}: Based on our findings, we conclude that splitting long-running tests can be an important measure for practitioners to cope with test flakiness. Test splitting enables parallelization of test executions and also reduces the cost of re-executions after flaky failures because the scope of the re-executed tests is narrower. % to the actual point of failure.
Thus, splitting long-running tests into smaller tests with a narrower scope can effectively decrease the negative effects of test flakiness in complex testing environments.
However, when splitting long-running tests, practitioners need to consider the potential test setup overhead of test splits.
\end{abstract}





\maketitle

\section{Lay Abstract}
%At SAP HANA, we found that all tests can be flaky, varying mainly in how often they fail. Flakiness interferes with continuous integration, leading to costly re-executions of flaky failures. 

%We analyzed test data from 12 weeks to examine how various test and environment characteristics affect the frequency of flaky failures. Then, we develop strategies to minimize the impact of flaky tests by identifying the factors that contribute to test flakiness. 

%We found the execution time of tests have a relatively large impact on test flakiness, which is also supported by previous research. The increased flakiness can either result from a test’s broader scope or slower execution from executing tests on slower infrastructure. 

%Interestingly, tests that need a distributed setup showed lower flakiness rates compared to non-distributed tests. We hypothesize that developers pay particular attention to designing distributed tests due to their inherent importance for SAP HANA.

%We found no significant correlation between the load on SAP HANA’s testing infrastructure and test flakiness. This might indicate effective load balancing. 

%The relationship between a test’s flakiness and memory and CPU requirements is inconclusive. One dataset showed a significant positive correlation, while another did not.

%One practical measure to reduce the negative impact test flakiness can be splitting long-running tests. This approach allows more tests to run in parallel and also lowers the cost of re-executions by isolating the point of failure. 

Test flakiness, that is, tests failing unpredictably without changes to the code, is a significant issue in software development. It disrupts the continuous integration (CI) process, making it more difficult for teams to maintain their software systems. Some researchers suggest that all tests experience flakiness, not necessarily due to the code base, but also due to problems in the testing environment.
In our study, we explore factors that might cause tests to become flaky, with the aim of helping software engineers develop strategies to reduce these random failures. We analyze two datasets derived from SAP HANA test data for a period of 12 weeks. One dataset is based on test results in the production environment, while the other is based on a special experiment designed to study flakiness.
Our findings showed that tests with longer execution times were more likely to be flaky. This might be because longer tests cover more functionality or are more affected by issues in the testing infrastructure. Interestingly, distributed tests were less flaky than those running on a single machine. We discovered that the load on the testing infrastructure did not seem to influence test flakiness, and the relationship between the memory and CPU needed for a test to run and the test's flakiness was unclear.
Based on these results, we recommend splitting longer tests into smaller ones. This approach not only allows for parallel test execution but also makes it easier to identify and fix flaky failures. However, it is important to balance the benefits of smaller tests against potential overheads.

\section{Introduction}
Test flakiness is a major problem in the software industry. Flaky tests yield different results when executed multiple times on the same code version. Thus, flaky tests impede continuous integration, because, to automatically merge a proposed code change~\cite{vocabulary-hana}, all tests need to pass. Since software engineers aim to achieve shorter and shorter release cycles, test flakiness has gained increasing attention from companies such as Google, Apple, Microsoft, Meta, and SAP~\cite{google, apple, microsoft, meta, SAP}.

A common strategy to mitigate the negative effects of test flakiness is to re-execute failing tests multiple times on the same code version. When the test passes in one of the re-executions, the initial failure is considered a \emph{flaky failure} and the test is viewed as passing~\cite{fallahzadeh2022impact, SAP, bach2018effects, google}. However, this strategy is costly with respect to computational resources. For example, for SAP HANA's pre-submit testing in the main code line alone, 500 hours of computing time daily are dedicated to restarts caused by flaky failures. Google reports that they use up to one day of computing time for every week they spend testing~\cite{micco2017state}.

Previous work highlighted that the probability of flaky failures increases with the complexity of both the executed tests and their execution environment~\cite{SAP, meta, durieux2020empirical}. For example, when tests are executed in distributed testing environments with heterogeneous hosts~\cite{micco2017state, SAP}, the number of flaky failures due to timeout flakiness might increase, as the increased variance of test execution times makes it harder for developers to determine appropriate timeout values~\cite{SAP}. When it comes to tests themselves, system tests covering a large scope with a low degree of isolation may exhibit a higher flakiness rate than smaller unit tests. However, system tests are particularly common in large-scale industrial software systems~\cite{meta, SAP}. 

The connection between test complexity and flakiness appears to be a consensus among practitioners~\cite{google-flaky, meta, apple}. Previous research found that characteristics related to test complexity are effective features for predicting whether a test is flaky~\cite{alshammari2021flakeflagger}. Furthermore, a Google blog post reported a strong correlation between the binary size and memory usage of a test and its flakiness rate~\cite{google-flaky}. 

In this paper, we present a study of test flakiness in the context of a large industrial database management system, SAP HANA. First, we validate previous findings on correlations between test characteristics and flakiness. We perform a correlation analysis between the flakiness rate of a test and its average execution time, the required number of CPU threads, and the required main memory. 

Based on internal discussions with practitioners at SAP, we further analyze whether tests that verify functionality in distributed environments exhibit higher flakiness rates than non-distributed tests. The assumption that distributed tests are more flaky appears intuitive because tests in distributed environments might be more susceptible to common flakiness categories such as \emph{Async wait}, \emph{Concurrency}, or \emph{Network}~\cite{parry2021survey, meta}. However, to our knowledge, there is no empirical evidence that this relationship exists.

In addition to the above test characteristics, we correlate two characteristics of SAP HANA's test environment with test flakiness. First, we validate previous findings that indicate a correlation between load on the testing system and flakiness~\cite{cordeiro2021shaker, silva2023effects, leinen2024impact}. Second, we examine the correlation between flakiness and the performance of the test execution host. More specifically, since SAP HANA's testing infrastructure is scaled out across more than \num{1000} heterogeneous physical hosts, we hypothesize that the flakiness rate increases on hosts with lower computational power.

To study the test and environmental characteristics mentioned above, we constructed two datasets, one arising from the productive CI environment of SAP HANA, and one from dedicated experiments to study test flakiness. In total, we obtained more than 1.5 million test results over a 12-week period. 

The main contributions of the study presented in this paper are: 

\begin{enumerate}
    \item Validation of previous findings on correlations between test (environment) characteristics and test flakiness using data from system tests of a large industrial database management system~\cite{silva2023effects, leinen2024impact, cordeiro2021shaker, alshammari2021flakeflagger, lampel2021life}.
    \item A new set of test (environment) characteristics, derived from discussions with practitioners at SAP, and their correlation with test flakiness.
    \item An evaluation of the practical usefulness of our results for a large-scale industrial project.
\end{enumerate}

In the following, we describe our study subject, the constructed datasets, our definition of test flakiness, and our research questions. Then, we present our results, discuss them, and conclude the paper with a discussion of threats to validity and concluding remarks.

%In \Cref{sec:background}, we describe our study subject, the constructed datasets, our definition of test flakiness, and our research questions.
%In \Cref{sec:results}, we report our empirical findings. We conclude the paper with a discussion of our findings (\Cref{sec:discussion}), threats to validity (\Cref{sec:threats}), and concluding remarks (\Cref{sec:conclusion}).

\begin{figure*}
     \centering
     \begin{subfigure}[b]{0.62\textwidth}
        \centering
        \includegraphics[width=\textwidth]{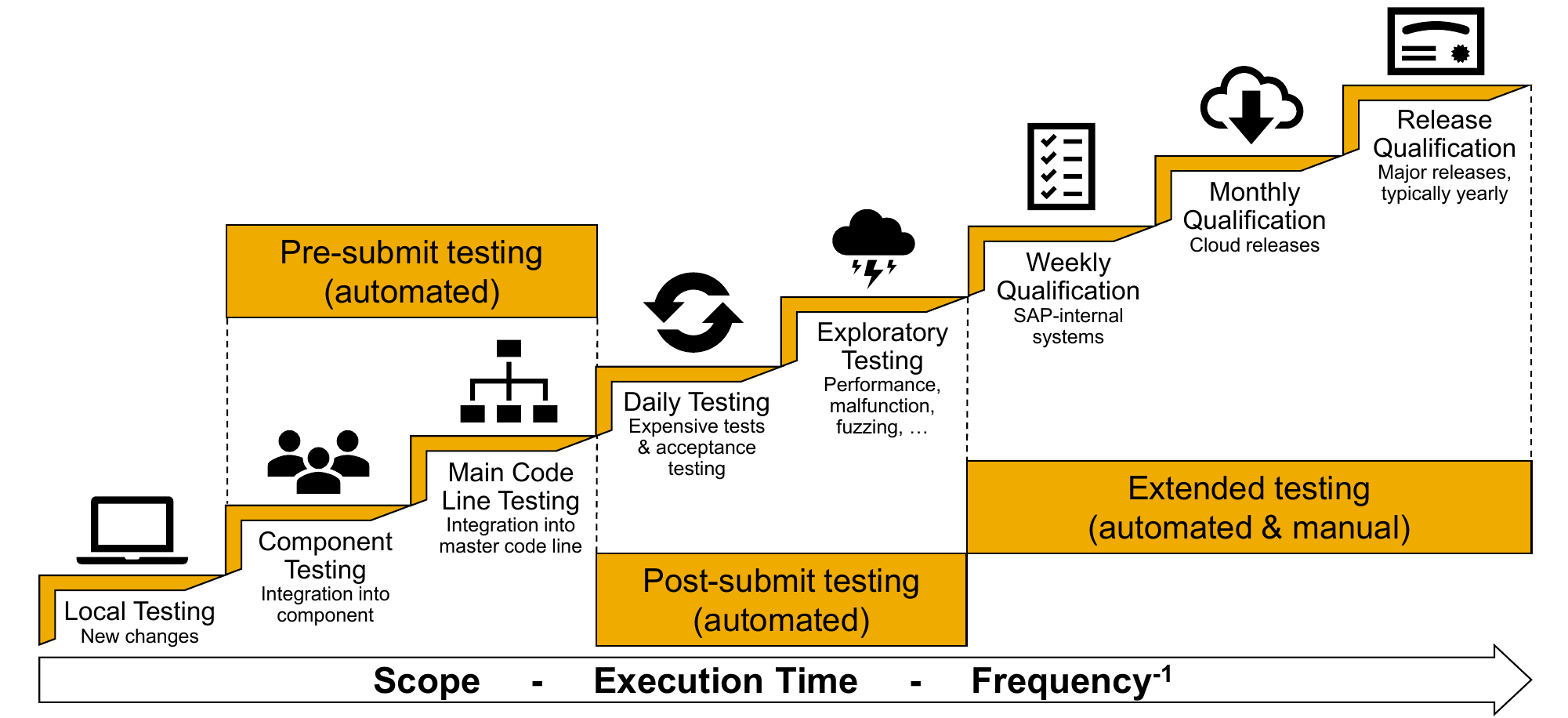}
        \caption{Testing stages of SAP HANA.}
        \Description[Testing stages of SAP HANA.]
        {Testing stages of SAP HANA.}
        \label{fig:testingstages}
     \end{subfigure}
     \hfill
     \begin{subfigure}[b]{0.36\textwidth}
        \centering
        \includegraphics[height=5.2cm]{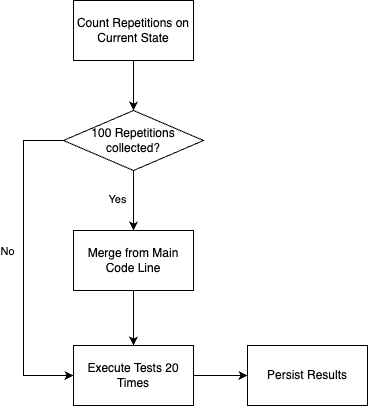}
        \caption{Data collection process for MT dataset~\cite{SAP}.}
        \Description[Data collection process for MT dataset]
        {Data collection process for MT dataset}
        \label{fig:mte-process}
     \end{subfigure}
    \caption{Study subject and data collection.}
    \label{fig:hana-figures}
\end{figure*}

\section{Background}
\label{sec:background}
In this section, we present the study background. First, we introduce the study subject, Second, we present the datasets used. Third, we present the research questions and the methods to answer them.

\subsection{Study Subject}
\label{sec:subject}

SAP HANA, the subject of this study, is a large-scale in-memory database management system that has been developed by SAP for more than ten years~\cite{bach2022testing}. To enable its use in business-critical customer scenarios, SAP HANA is extensively tested in multiple stages that vary in scope and execution frequency, as shown in \Cref{fig:testingstages}. Bach et al. provide a more detailed overview of testing at SAP HANA in~\cite{bach2022testing}.

This work focuses on system tests in the so-called pre-submit testing stage, i.e., the tests that are executed before a developer's change is merged into the main code line.  
Since executing all pre-submit tests sequentially would take more than three days, the tests are distributed over multiple hosts and executed in parallel. For parallelization, the tests are grouped according to their required configuration of SAP HANA. In a test run, each group of tests is then executed against a dedicated installation of SAP HANA running in a Docker container. In many cases, these tests contain SQL statements to communicate with the system under test~\cite{bach2022testing}. 
In this study, we focus on the system tests in SAP HANA's pre-submit stage. Thus, the scope of our study is approximately 800 tests that account for more than 90\% of the test resource consumption in the SAP HANA pre-submit testing stage.

To cope with the immense computational demands of testing, SAP HANA's testing infrastructure is scaled out across approximately \num{1000} physical hosts that vary in their age, location, number of CPU threads, memory, and processor generation~\cite{bach2022testing}. For every change to be merged, the executed tests are distributed on more than a hundred physical hosts. Furthermore, since SAP HANA can also be used in a distributed setup, some tests require multiple hosts on their own. 
As distributed testing systems are commonly used for large software projects in the software industry~\cite{google, meta}, we analyze potential environmental characteristics that could lead to test flakiness, with the goal of better understanding the impact of the test environment on flaky failures. 

\subsection{Datasets}
For SAP HANA, the results of all test executions are stored in a database, together with metadata about the tests and the execution hosts. In this study, we utilize this data to construct two different datasets.

First, in the \emph{Production Environment} dataset (PE), we collect production data from SAP HANA's CI pipeline. That is, we collect the results of system tests executed in SAP HANA's pre-submit testing stage within a 12-week period. As a result, we end up with more than 1.5 million test executions.

Second, we collect data from our so-called \emph{Mass Testing} (MT) experiment~\cite{SAP}. The process of the experiment is illustrated in \Cref{fig:mte-process}. The idea of the mass testing experiment is to repeatedly execute SAP HANA's pre-submit tests on the same version of the code to study test flakiness. To create the dataset, we used idle resources in the testing infrastructure on weekends. 
Our objective is to aggregate 100 repeated test executions per code version. To achieve this, we run a job that executes SAP HANA's pre-submit tests 20 times for the same code version every weekend. The job first checks if there already exist 100 test repetitions with the current version of the code. If so, the code is updated to the most recent version of the main code line. If there are less than 100 repetitions, the job executes the tests to collect data from 20 additional repetitions.
Based on this process, we have collected more than \num{100000} test executions in the same 12-week period as for the PE dataset. We provide an overview of the datasets in \Cref{tab:data-sets}.

\begin{table}
  \begin{center}
  \caption{Available datasets for this study.}
    \begin{tabular}[width=\textwidth / 2]{l r r}
    \toprule
      \textbf{Dataset} & \textbf{\# Tests} & \textbf{\# Test Executions}\\
      \midrule
      PE & 721 & \num{1528986}\\
      MT & 688 & \num{129660}\\
      \bottomrule
      \end{tabular}
    \label{tab:data-sets}
  \end{center}
\end{table}

\subsection{Flakiness Definition}
\label{sec:definition}

The traditional informal definition of flakiness classifies a test as flaky when it yields different results for repeated executions on the same version of the code~\cite{parry2021survey}. Previous research examined different features associated with this binary notion of flakiness to predict whether a test is flaky~\cite{fatima2022flakify, vocabulary-hana, pinto2020vocabulary, bell2018deflaker, lam2019idflakies, pontillo2022static}. However, previous research also pointed out that, especially in large testing environments, quantifying flakiness beyond this binary notion can be useful, as all tests exhibit some level of flakiness~\cite{apple}.
In a previous study at SAP HANA, we found that most system tests in SAP HANA's pre-submit test suite show some degree of flakiness when the number of test repetitions grows towards infinity~\cite{SAP}. Therefore, for this work, rather than relying on the binary notion of flakiness, we examine the failure rate of flaky tests~\cite{SAP, silva2023effects}. 

More formally, given a test $t$ we compute $\mathcal{R}_p(t)$, the \emph{flakiness rate} of $t$ in period $p$, by dividing the number of flaky executions $f_p(t)$ by the total number of executions $e_p(t)$ of that test $t$ in period $p$. Here, we label an execution as flaky if the test failed on a certain version of the code for which it also showed passing results.
Thus, the formula for calculating the flakiness rate $\mathcal{R}_p \in [0,1]$ of $t$ in $p$ is:
\begin{equation*}
\mathcal{R}_p(t) = \frac{f_p(t)}{e_p(t)}.
\end{equation*}

This approach allows us to quantify the effect that different characteristics have on the frequency of flaky failures beyond a binary flakiness definition. 
It further allows us to estimate the additional computational cost caused by flaky tests because the costs of re-executions increase with the frequency of flaky failures~\cite{SAP, apple}.

\subsection{Research Questions and Methods}
\label{sec:questions}

Our study is based on two main research questions.
In the following, we introduce and motivate these questions and outline the methods we use to answer them.

\begin{quote}
    \textbf{RQ1}: How do different test complexity characteristics correlate with test flakiness in the context of SAP HANA?
\end{quote}

Previous research indicates that higher test complexity can lead to a higher flakiness rate~\cite{SAP, meta, alshammari2021flakeflagger, apple}. To answer \textbf{RQ1}, we evaluate various test characteristics related to test complexity to check whether these characteristics have a significant effect on the flakiness rate of tests in the context of SAP HANA. Some of our hypotheses are derived from related work, while others are the result of internal discussions with practitioners at SAP.
\Cref{tab:research-questions} summarizes the characteristics that we examined and, if available, the previous work that motivates them. In the following, we discuss the motivation for the selected characteristics in detail. 

\begin{table}
    \centering
    \caption{Test and environment characteristics: We hypothesize that each of these characteristics has an impact on the flakiness rate of tests in the context of SAP HANA.}
    \begin{tabular}{l l c}
         \hline
         \textbf{Category} & \textbf{Characteristic} & \textbf{Related work} \\
         \hline
         Test & Test execution time & \cite{alshammari2021flakeflagger, lampel2021life}\\
         & CPU & \cite{silva2023effects, leinen2024impact, cordeiro2021shaker}\\
         & Memory & \cite{silva2023effects}\\
         & Distributed test & None \\
         \hline
         Test environment & Host performance & None\\ % Execution host performance
         & System load & \cite{cordeiro2021shaker, silva2023effects, leinen2024impact}\\
         \hline
    \end{tabular}
    \label{tab:research-questions}
\end{table}

\textbf{On test execution time}:
Previous research has found that test execution time can be an effective feature to predict whether a test is flaky~\cite{alshammari2021flakeflagger}. However, the reasons for flakiness have been shown to vary between different types of software~\cite{gruber2022survey} and between programming languages~\cite{barbosa2022test}. Therefore, in this study, we examine the relationship between test execution time and flakiness rate in the context of SAP HANA, a large-scale database management system written mainly in C++.

We first validate the findings of previous research suggesting that test execution time is correlated with test flakiness~\cite{alshammari2021flakeflagger}. We further quantify the strength of this relationship with the help of \emph{Pearson's correlation coefficient} $r$ and report the corresponding $p$-value  with a significance threshold of $p \leq 0.05$.
%Finally, to verify the statistical significance of the relationship, we perform \emph{Pearson's r test} with a significance threshold of $p \leq 0.05$.

\textbf{On required resources}:
An article in Google's testing blog states that the memory usage of a test explains part of the variance in its flakiness rate (r2=0.76)~\cite{google-flaky}. The article further reports that tests labeled \enquote{large} by developers show a higher flakiness rate than tests labeled to be of \enquote{medium} size, which, in turn, are more likely to be flaky than tests labeled as \enquote{small}~\cite{google-flaky}.
In the SAP HANA organization, developers must provide detailed metadata on the resources required for running a test. As described in \Cref{sec:subject}, for each group of tests, developers assign a certain host configuration that contains the number of available CPU threads and memory.  

Based on the findings reported in the Google article and previous research that found a relationship between flakiness and CPU/memory constraints~\cite{leinen2024impact, silva2023effects, cordeiro2021shaker}, we hypothesize that tests requiring more computational resources in terms of CPU or memory show a higher flakiness rate. 
To validate this hypothesis, we group the tests based on their required resources as labeled by SAP HANA's developers and check whether higher memory or CPU requirements correlate with the tests' flakiness rates. 

\textbf{On distributed environments}:
As mentioned in \Cref{sec:subject}, some tests in SAP HANA's pre-submit stage require a distributed setup because they verify distributed functionality (e.g., distributed transactions~\cite{farber2012sap}).
As motivated above, tests in distributed environments might be more susceptible to common flakiness categories.
%According to the CAP theorem, consistency in distributed systems requires a trade-off with the availability and partition tolerance of the system~\cite{cap}. 
Therefore, we hypothesize that such \emph{distributed tests} show a higher flakiness rate than non-distributed tests.

\begin{quote}
    \textbf{RQ2}: How do different test environment characteristics correlate with test flakiness in the context of SAP HANA?
\end{quote}

Previous research claims that more complex testing environments can lead to more flakiness~\cite{SAP, meta, apple}. In this research question, our aim is to verify this claim by breaking the complexity of SAP HANA's testing environment down into concrete characteristics of the environment as listed in \Cref{tab:research-questions}. To answer \textbf{RQ2}, we use a similar method as for \textbf{RQ1}: we check whether environmental characteristics show a significant correlation with the flakiness rate of SAP HANA's system tests. In the following, we motivate our selection of environmental characteristics. 

\textbf{On execution host performance}:
As mentioned in \Cref{sec:subject}, SAP HANA's testing infrastructure consists of approximately \num{1000} heterogeneous hosts, i.e., physical servers located in SAP's data centers. These hosts were purchased in waves within the last ten years and vary in terms of their computational performance, for example, due to differences in the built-in processor generation.
Based on our assumption that the flakiness rate of a test increases with test execution time, as described for \textbf{RQ1}, we hypothesize that the flakiness rate on a host decreases with the host's performance.

\textbf{On system load}:
Previous research has shown that tests might fail flakily due to resource limits~\cite{leinen2024impact, silva2023effects}. We further investigate the idea of resource limits impacting flakiness by assessing the relationship between system load and test flakiness. 
To achieve this, we operationalize the load on our testing infrastructure as the number of tests executed per hour.
We then check whether this notion of system load correlates with the flakiness rate of the executed tests.

\section{Empirical Results}
\label{sec:results}

In this section, we present our results along the research questions and the characteristics motivated above. 

\subsection{RQ1: Test Complexity}
\label{sec:rq1}
To answer \textbf{RQ1}, we evaluate different test complexity characteristics and their correlation with the flakiness rate of SAP HANA's system tests. In the following, we present our results for each of the characteristics listed in \Cref{sec:questions}.

\subsubsection{Average test execution time}
Based on the assumption that longer-running tests yield higher flakiness rates, we calculate the average execution time for each of SAP HANA's system tests. 
To this end, we first exclude outliers by removing test executions with execution times that are outside the $10th$ to $90th$ percentile in the period considered. We then calculate the arithmetic mean of the remaining execution times and examine their relationship with the flakiness rate of the respective test as defined in \Cref{sec:definition}.

To ensure a reasonable sample size of test executions per test for our analysis, we remove tests from the PE dataset that were executed less than 50 times in the given period. In the filtered dataset, the average number of executions per test is \num{1752}. 

To further reduce the impact of outliers in our dataset, we divide the tests into equal-width bins based on their execution time, i.e., every bin contains tests for which the average execution time lies within a 5-minute interval. Since the execution times range from a few seconds to almost one hour, we end up with 12 bins. \Cref{tab:time-groups} shows the resulting bins and the number of tests per bin. 

\begin{figure}
    \centering
    \includegraphics[width=\linewidth]{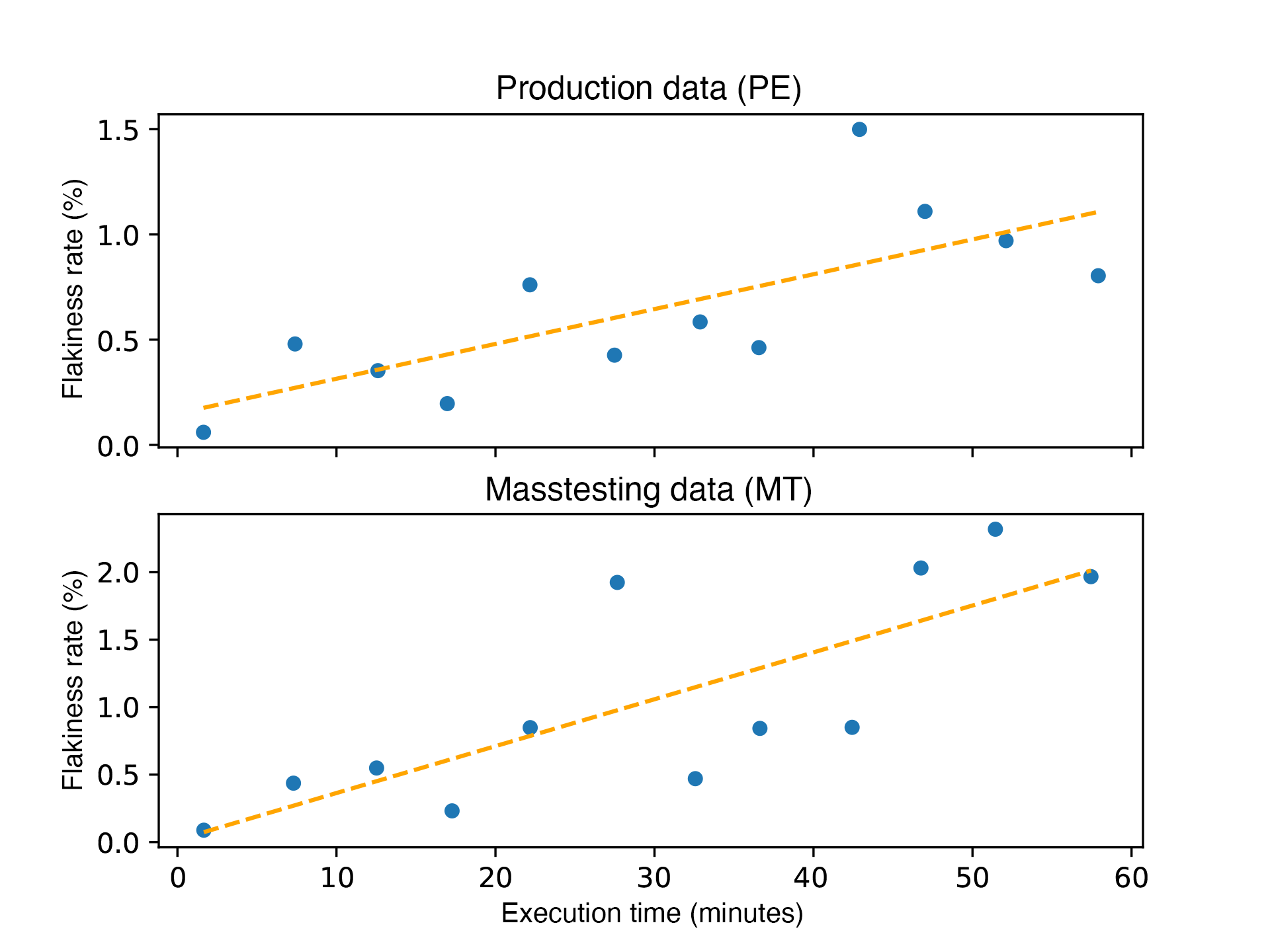}
    \caption{Scatterplot showing the relation between test execution time and mean flakiness rate. The orange line depicts the fitted regression line.}
    \Description[Scatterplot showing the relation between test execution time and mean flakiness rate.]{Scatterplot showing the relation between test execution time and mean flakiness rate.}
    \label{fig:execution-times}
\end{figure}

\Cref{fig:execution-times} visualizes the relationship between average test execution time and flakiness rate. The orange line represents the regression line, which we obtain by fitting a \emph{linear regression model}. To quantify the strength of the relationship, we calculate \emph{Pearson's correlation coefficient} $r$ together with the corresponding $p$-value using the \texttt{pearsonr} function from the SciPy library with default settings~\cite{scipy-docs}.
We report the $r$ values for the two datasets separately using indices (e.g., $r_{PE}$ is the correlation coefficient in the PE dataset).

As shown in \Cref{tab:rq1}, the correlation coefficients for the two datasets are $r_{PE} = 0.74$ and $r_{MT} = 0.79$, suggesting that there exists a positive correlation between average test execution time and flakiness rate. As the resulting p-values as shown in \Cref{tab:rq1} are below our threshold of $0.05$, we conclude that the flakiness rate is indeed correlated with the average test execution time. 

\begin{table}
  \begin{center}
  \caption{Pearson's correlation coefficient ($r$) of average test execution time and flakiness rate together with $p$-values.}
    \begin{tabular}[width=\textwidth / 2]{l r r}
    \toprule
      \textbf{Dataset} & \textbf{$r$} & \textbf{$p$-value}\\
      \midrule
      PE & 0.74 & 0.0063\\
      MT & 0.79 & 0.0022\\
      \bottomrule
      \end{tabular}
    \label{tab:rq1}
  \end{center}
\end{table}

\framed{Answer RQ1 (execution time)}{There exists a significant positive correlation between average test execution time and flakiness rate in both datasets. The correlation coefficient of this relationship is 0.74 for the PE and 0.79 for the MT dataset.}

\begin{table}
  \begin{center}
  \caption{Equal-width execution time bins and the respective number of tests per bin for the two datasets.}
    \begin{tabular}[width=\textwidth / 2]{l r r}
    \toprule
      \textbf{Interval} & \textbf{\# Tests PE} & \textbf{\# Tests MT}\\
      \midrule
      (0, 5] & 376 & 361\\
      (5, 10] & 90 & 88\\
      (10, 15] & 72 & 65\\
      (15, 20] & 55 & 57\\
      (20, 25] & 32 & 23\\
      (25, 30] & 18 & 24\\
      (30, 35] & 22 & 22\\
      (35, 40] & 16 & 14\\
      (40, 45] & 11 & 17\\
      (45, 50] & 7 & 7\\
      (50, 55] & 3 & 4\\
      (55, 60] & 4 & 4\\
      \bottomrule
      \end{tabular}
    \label{tab:time-groups}
  \end{center}
\end{table}

\subsubsection{Required resources}
In the following, we investigate the relationship between a test's flakiness rate and its resource requirements as labeled by SAP HANA's developers. As mentioned in \Cref{sec:subject}, developers assign their tests to a certain pool of hosts that satisfy the required computational resources. Every pool is defined by the available CPU threads and main memory per host. We use these labels to investigate whether \enquote{larger} tests show higher flakiness rates. 
% As shown in \Cref{fig:pools}, there exist a variety of different CPU / memory combinations. Overall, the PE datasets contains 23 different combinations, the MT dataset 22. The available resources in the pools range from 4 to 32 CPU threads and 16GB to 128GB memory.

For our analysis, we group the tests in the two datasets according to the assigned memory and CPU requirements, respectively. Based on the resulting groups, we calculate the flakiness rate of each group and test whether there exists a positive correlation between the flakiness rate and the respective memory or CPU labels. 

\begin{figure}
    \centering
    \includegraphics[width=\linewidth,trim={0 0pt 0 0pt},clip]{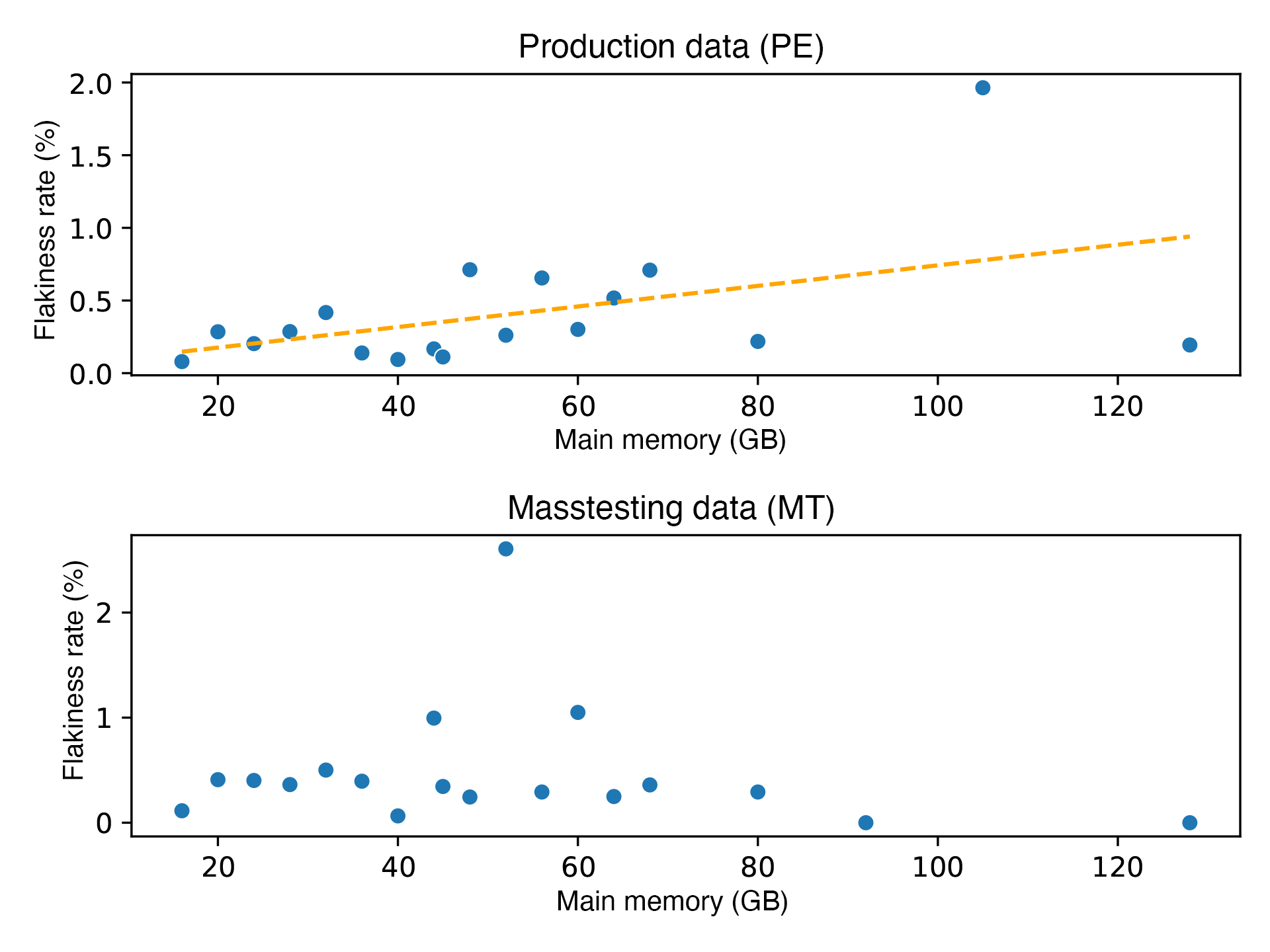}  % [trim={left bottom right top}
    \caption{Scatterplot showing the relation between available memory and mean flakiness rate. Significant positive correlation in PE dataset, no significant correlation in MT dataset. }
    \Description[Scatterplot showing the relation between available memory and mean flakiness rate.]{Scatterplot showing the relation between available memory and mean flakiness rate. Significant positive correlation in PE dataset, no significant correlation in MT dataset.}
    \label{fig:memory}
\end{figure}

\Cref{fig:memory} shows the average test flakiness rates in relation to the assigned memory labels for the PE and MT datasets. We find that there exists a significant positive correlation between assigned memory label and flakiness rate in the PE dataset ($r_{PE} = 0.47$, $p = 0.05$). In the MT dataset, however, we observe no significant correlation between flakiness rate and assigned memory label ($r_{MT} = -0.14$, $p = 0.6$). 

\begin{figure}
    \centering
    \includegraphics[width=\linewidth,trim={0 0pt 0 0pt},clip]{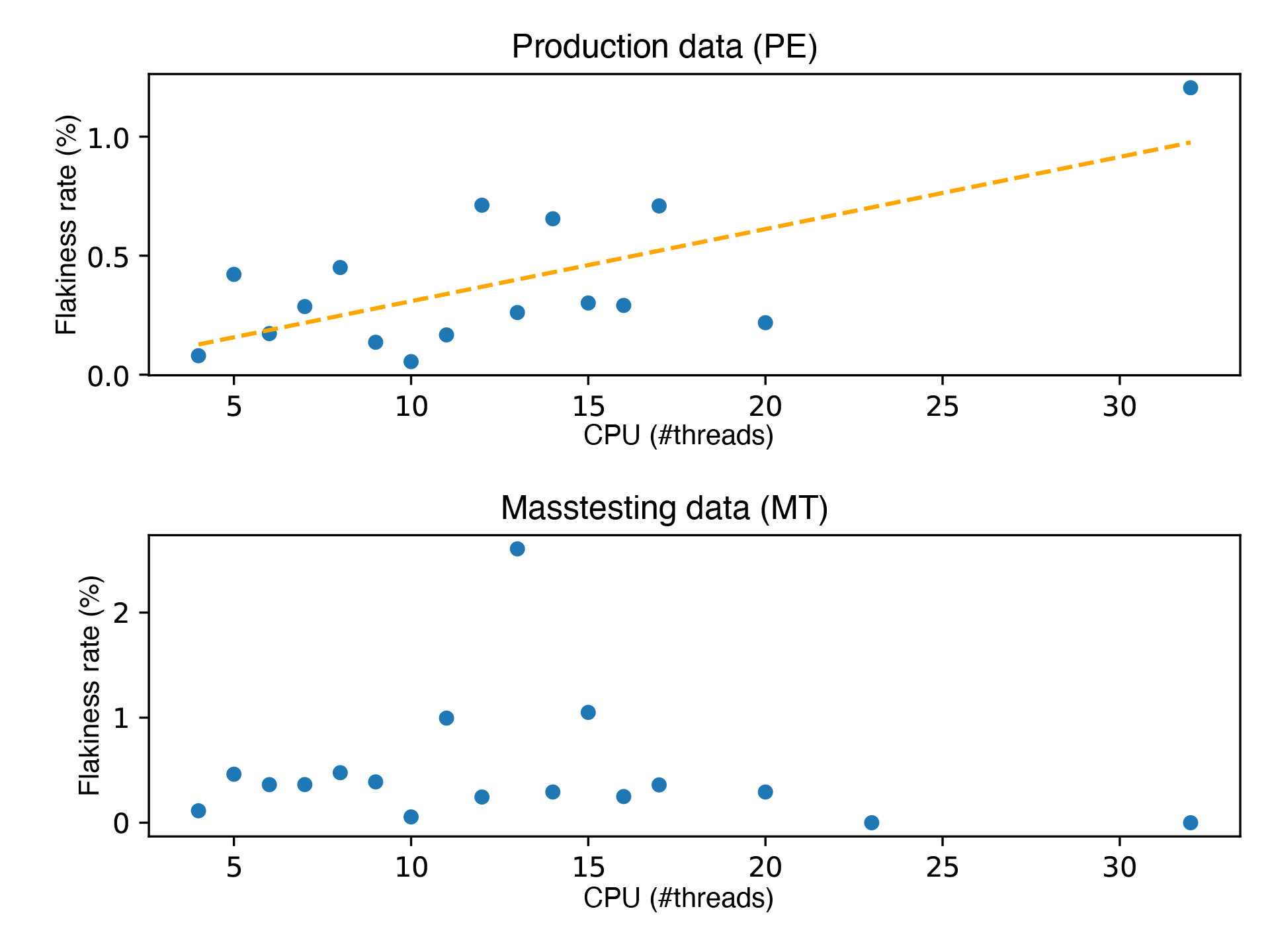}  % [trim={left bottom right top}
    \caption{Scatterplot showing relation between available CPU threads and mean flakiness rate. Significant positive correlation in PE dataset, no significant correlation in MT dataset.}
    \label{fig:cpu}
    \Description[Scatterplot showing relation between available CPU threads and mean flakiness rate.]{Scatterplot showing relation between available CPU threads and mean flakiness rate. Significant positive correlation in PE dataset, no significant correlation in MT dataset.}
\end{figure}

As visible in \Cref{fig:cpu}, these findings also apply to the assigned CPU labels. Although there exists a significant positive correlation between assigned CPU labels and flakiness rates in the PE dataset ($r_{PE} = 0.69$, $p = 0.03$), we do not observe a significant relationship in the MT dataset ($r_{MT} = -0.13$, $p = 0.6$).

\framed{Answer RQ1 (resources)}{The relationship between the flakiness rate and the required test resources is inconclusive. Although there exist significant correlations between the flakiness rate and both the number of assigned CPU threads and the assigned main memory in the PE dataset, we do not observe any significant correlations in the MT dataset.}

% \begin{figure}
    %\centering
    %\includegraphics[width=\linewidth]{images/resources.png}
    %\caption{Memory and CPU for the different pool sizes of the MT dataset}
%    \label{fig:pools}
% \end{figure}

\subsubsection{Distributed tests}
In the following, we investigate whether distributed tests are more flaky than tests that do not require distributed environments. 
As explained in \Cref{sec:subject}, all test executions and their results with additional metadata are stored in a database. The metadata about test executions contains a label on whether the test requires a distributed environment in its setup, which allows us to examine whether such distributed tests have an increased flakiness rate. 

For our analysis, we divide the tests into two groups, based on the binary label \enquote{is distributed}.
Overall, approximately 6 \% of SAP HANA's pre-submit tests in our study belong to the distributed group. We then calculate the average flakiness rate for both groups and compare them. Finally, we conduct a \emph{Mann-Whitney U test} as implemented by SciPy. The Mann-Whitney U test is a nonparametric test of the null hypothesis that the underlying distribution of two samples is the same~\cite{scipy-docs-mann}. In our case, we use Mann-Whitney U to determine whether the distribution of flakiness rates for distributed tests is significantly different from the flakiness rates for non-distributed tests~\cite{scipy-docs-mann}. 

\begin{figure}
    \centering
    \includegraphics[width=0.95\linewidth]{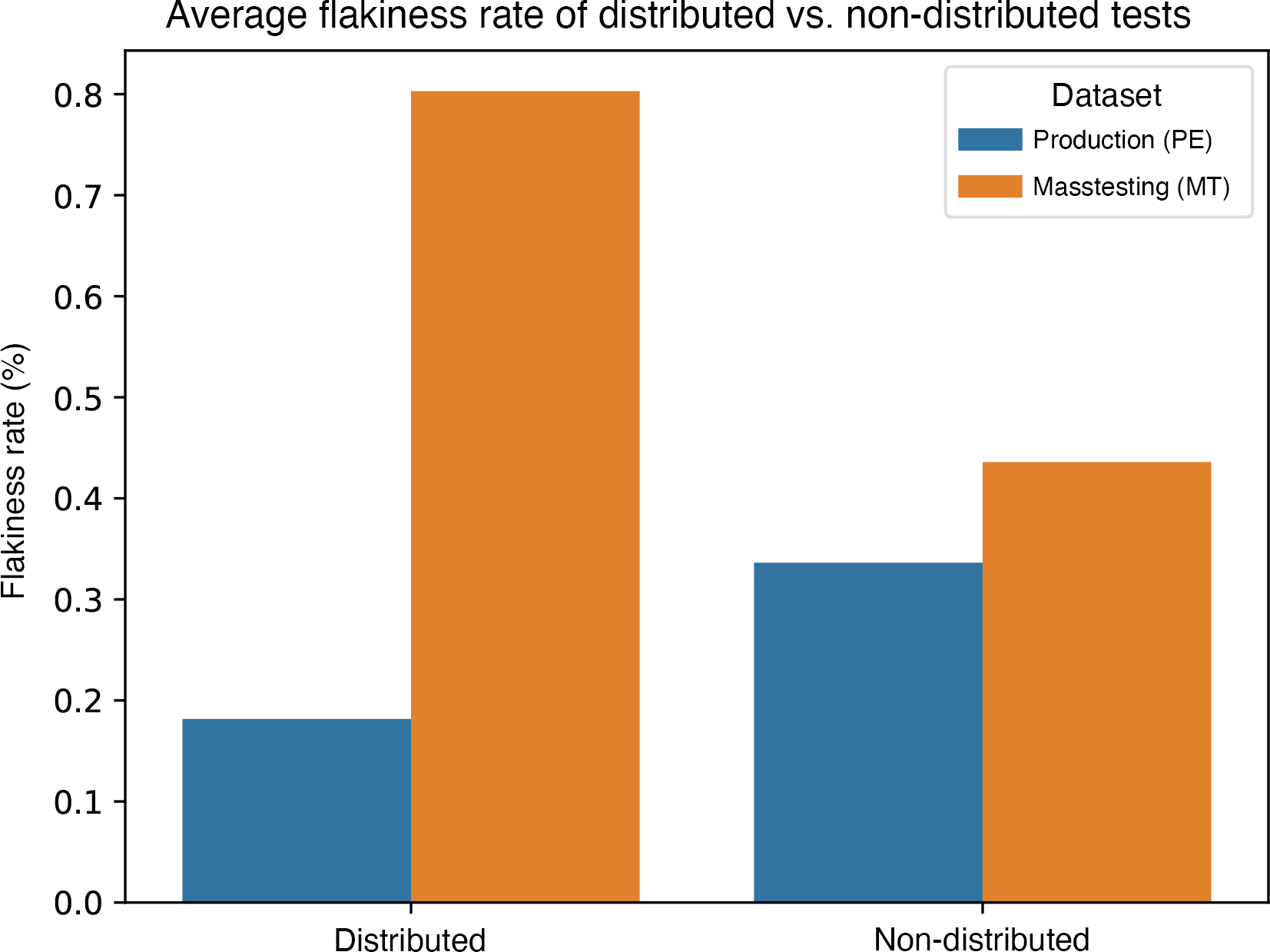}
    \caption{Arithmetic mean of test flakiness rate in both datasets, tests divided by the \enquote{is distributed}-label. Note that the mean for distributed tests on the MT dataset was heavily influenced by a single test with a flakiness rate of 17.1\%.}
    \label{fig:dist}
    \Description{A bar plot showing the arithmetic mean of the test flakiness rate in both datasets, tests divided by the \enquote{is distributed}-label.}
\end{figure}

As shown in \Cref{fig:dist}, the results differ between the two datasets. While distributed tests appear more flaky in the MT datasets, the opposite is true for the PE dataset. 
Looking at individual tests and their flakiness rates, we found that the distributed test with the highest flakiness rate is the same for both datasets. The test verifies distributed streaming functionality and faced an \emph{async wait} issue leading to flaky failures. However, this issue was fixed by a developer within the 12-week period we consider for this study. 
As we updated the code version in our mass testing experiment only after 100 test repetitions, i.e., every fifth week, the issue caused a high number of flaky failures and led to a flakiness rate of 17.1 \% in the MT dataset, which heavily influenced the mean of the distributed tests. Removing this test from the data decreases the mean flakiness rate for the distributed tests in MT from 0.8 to 0.4, which then leads to a result similar to that in the PE dataset. 

Looking at the results of our \emph{Mann-Whitney U test}, we find that the flakiness rate of distributed tests is significantly lower for distributed tests in the PE dataset ($p = 5.23 \times 10^{-13}$).
However, the difference is not significant in the MT dataset ($p = 0.97$).

\framed{Answer RQ1 (distributed tests)}{In the given datasets, we do not observe an increase in the flakiness rate when testing distributed environments. In contrast, we observe that distributed tests show a significantly lower flakiness rate than non-distributed tests in the PE dataset.}

\subsection{RQ2: Test Environment}
\label{sec:rq2}
To answer \textbf{RQ2}, we assess the correlation of several characteristics in SAP HANA's testing environment (see \Cref{sec:questions}) with the flakiness rate of system tests only based on production data (PE), because our experimental data (MT) does not contain enough information about the environment.
%In the following, we present our results for each of the characteristics listed in .
%To answer \textbf{RQ2}, we assess the correlation of several characteristics in SAP HANA's testing environment. For this, we only employ the PE dataset, as the MT dataset does not contain sufficient information on the environment. In the following, we present our results for each of the characteristics listed in \Cref{sec:questions}.

\subsubsection{Execution Host Performance}
First, we investigate the correlation between the execution host's computational performance and the flakiness rate.

As mentioned in \Cref{sec:subject}, SAP HANA's testing infrastructure consists of approximately \num{1000} hosts. These hosts are grouped into pools, depending on their available resources. In the scheduling phase, the tests are executed in Docker containers, which run on the respective hosts, depending on their pool label. These pool labels define CPU and main memory requirements, which are then granted to the respective Docker container. Due to this virtual layer handling resource assignments, each physical host can belong to $n$ pools, as long as it fulfills the required resources for the pool configuration. Furthermore, each pool is assigned $m$ hosts. 

Once a test run is started, the tests are scheduled on the different hosts by a load balancer to make sure that the load is evenly distributed over the infrastructure. As the hosts have different resource configurations, the set of tests that the hosts execute differs. Therefore, we cannot simply compare the average test execution times per host to assess whether they influence the flakiness rates. 

\begin{figure}
    \centering
    \includegraphics[width=0.95\linewidth]{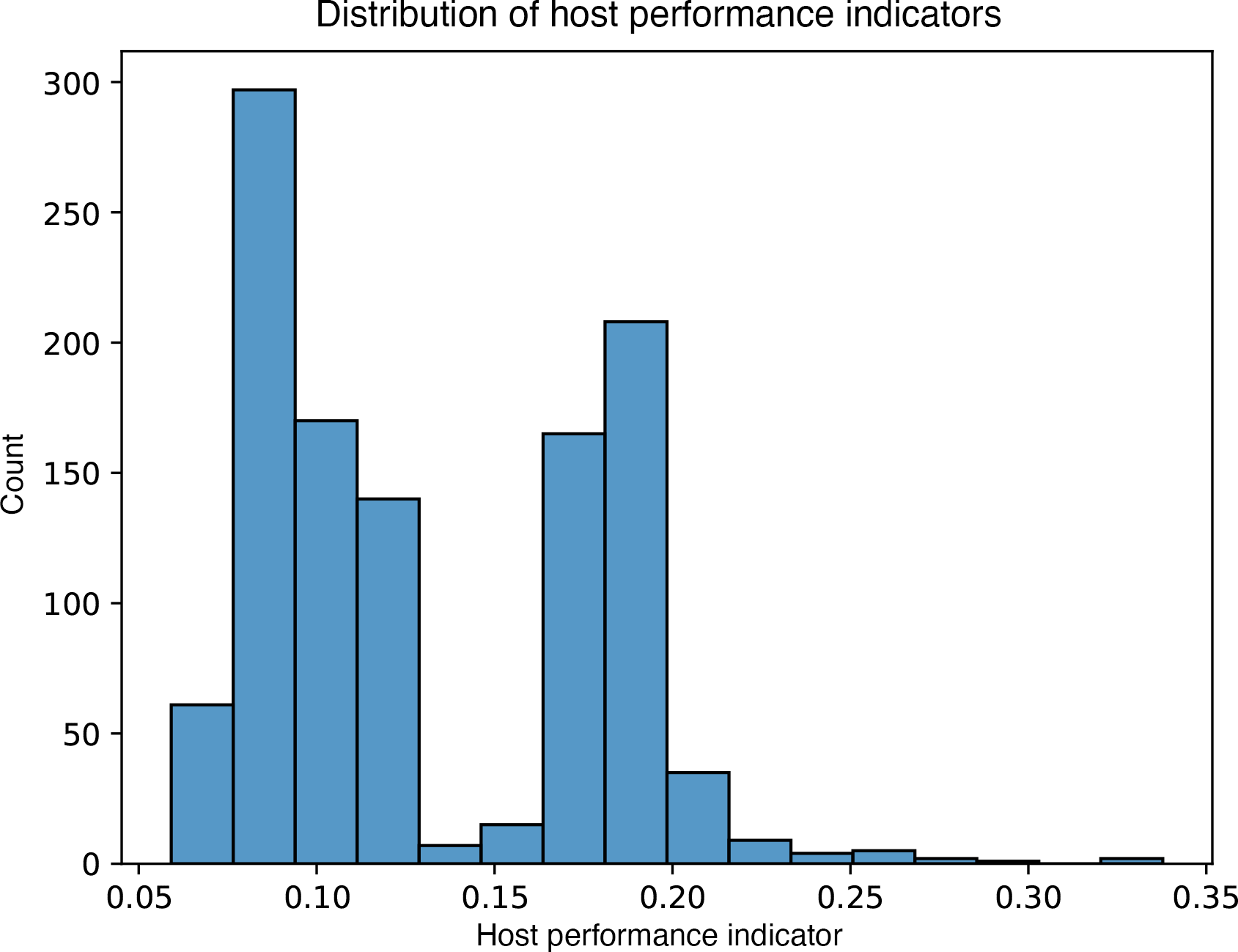}
    \caption{Histogram showing the distribution of host performance indicators obtained by averaging MinMax-scaled test execution times on a host.}
    \label{fig:hostperformance}
    \Description{Histogram showing the distribution of host performance indicators obtained by averaging MinMax-scaled test execution times on a host.}
\end{figure}

To mitigate this problem, we scale the execution times \emph{within-test} using sklearn's \emph{MinMaxScaler}~\cite{sklearn-minmax}. 
That is, let $E = \{e \in \mathbb{R}_0^+\}$ be the set of execution times of a certain test. After filtering out outliers by focusing on values between the 10th and 90th percentile, we scale each execution time $e \in E$ using the following formula: $\hat{e} = \frac{e - min(E)}{max(E) - min(E)} \in [0, 1]$. Thus, $\hat{E} = {\hat{e} \in [0, 1]}$ contains values in $[0, 1]$ for each test. We use the average of these scaled execution times as performance indicators for our hosts. Intuitively, the closer $\hat{e}$ for a given execution is to 1, the closer the execution time is to the maximum execution time of the respective test in $E$. 
\Cref{fig:hostperformance} shows the resulting distribution of host performance indicators, where a higher performance indicator implies that tests yield longer execution times on the respective host. 

To examine the relationship between host performance and flakiness rate, we group test executions based on the execution host and calculate average flakiness rates per host. According to our flakiness definition in \Cref{sec:definition}, we label an execution as flaky if it led to a failure and there exists an execution of the same test on the same version of the code that passed, although this passing run might have been executed on a different host. Similar to our approach in \Cref{sec:rq1}, we group hosts into 100 equal-width bins based on their performance indicator. 

\begin{figure}
    \centering
    \includegraphics[width=\linewidth, trim={0 0 0 10pt},clip]{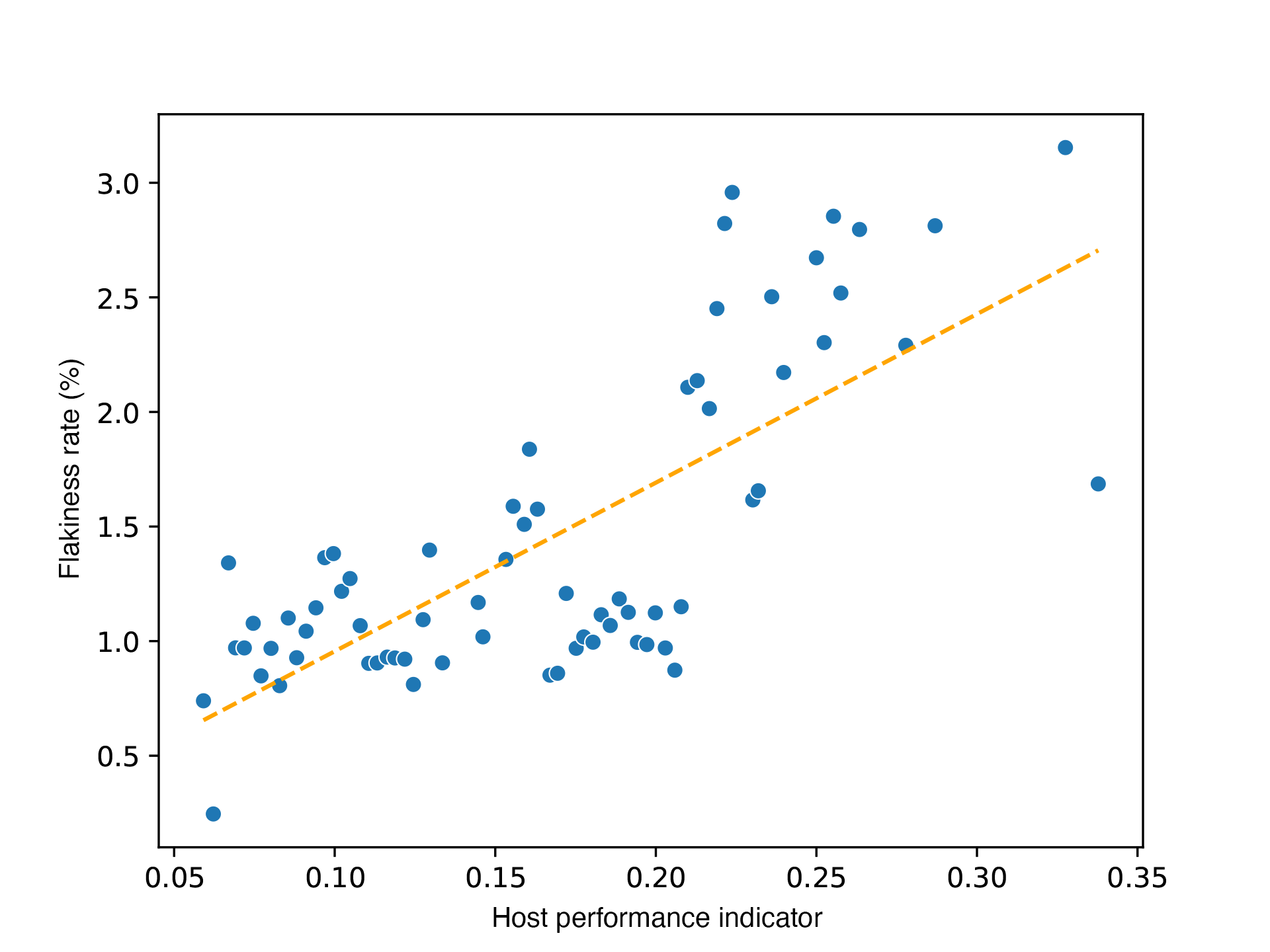}  % [trim={left bottom right top}
    \caption{Scatterplot showing the relation between host performance and mean flakiness rate. The orange line depicts the result of fitting a linear regression.} 
    \Description[Scatterplot showing the significant positive relation between host performance and mean flakiness rate.]{Scatterplot showing the relation between host performance and mean flakiness rate.}
    \label{fig:hostregression}
\end{figure}

\Cref{fig:hostregression} shows the relationship between the host performance indicator and the flakiness rate. We note that there exists a significant positive correlation ($r_{PE} = 0.74$). Since the $p$-value is below our threshold of $0.05$ ($p = 4.45 \times 10^{-13}$), we conclude that the flakiness rate increases significantly when the performance indicator of an execution host increases, i.e. when tests run longer on average on that host.

\framed{Answer RQ2 (host performance)}{There exists a significant positive correlation between the host performance indicator we defined (mean filtered and scaled test execution time per host) and the flakiness rate. The correlation coefficient for this relationship is 0.74 ($p < 0.05$), which indicates that the flakiness rate increases when the computational performance of the execution host decreases.}

\subsection{System Load}
In the following, we report the results of our correlation analysis on system load and flakiness rate. As mentioned in \cref{sec:questions}, we operationalize the load on our testing system as the number of tests that were executed per hour. 

\begin{figure*}
    \centering
    \begin{subfigure}[b]{0.49\textwidth}
        \centering
        \includegraphics[width=\textwidth]{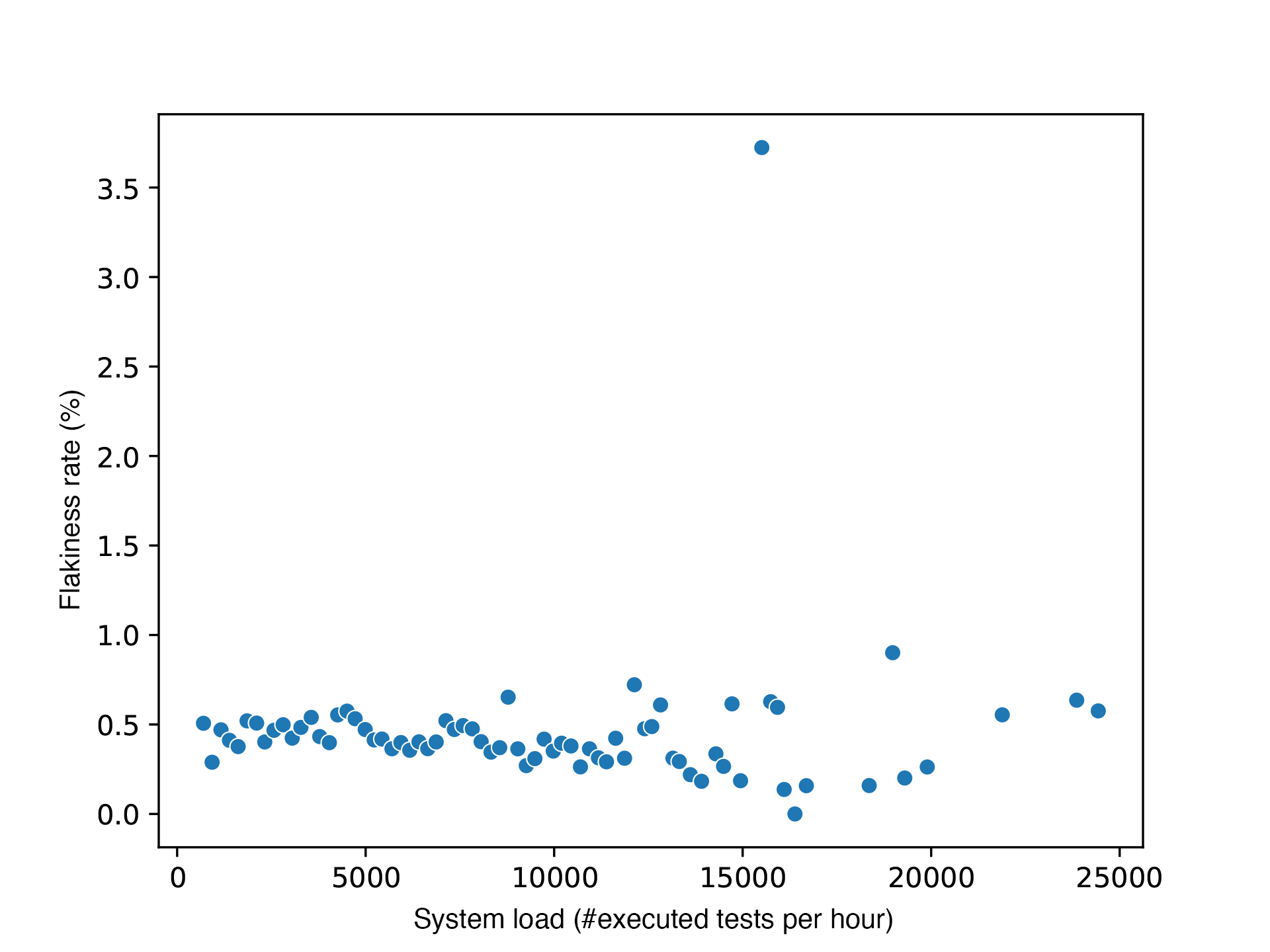}
        \caption{Scatterplot: Mean flakiness rate vs. system load.}
        \label{fig:load}
    \end{subfigure}
    \begin{subfigure}[b]{0.49\textwidth}
        \centering
        \includegraphics[width=0.95\textwidth]{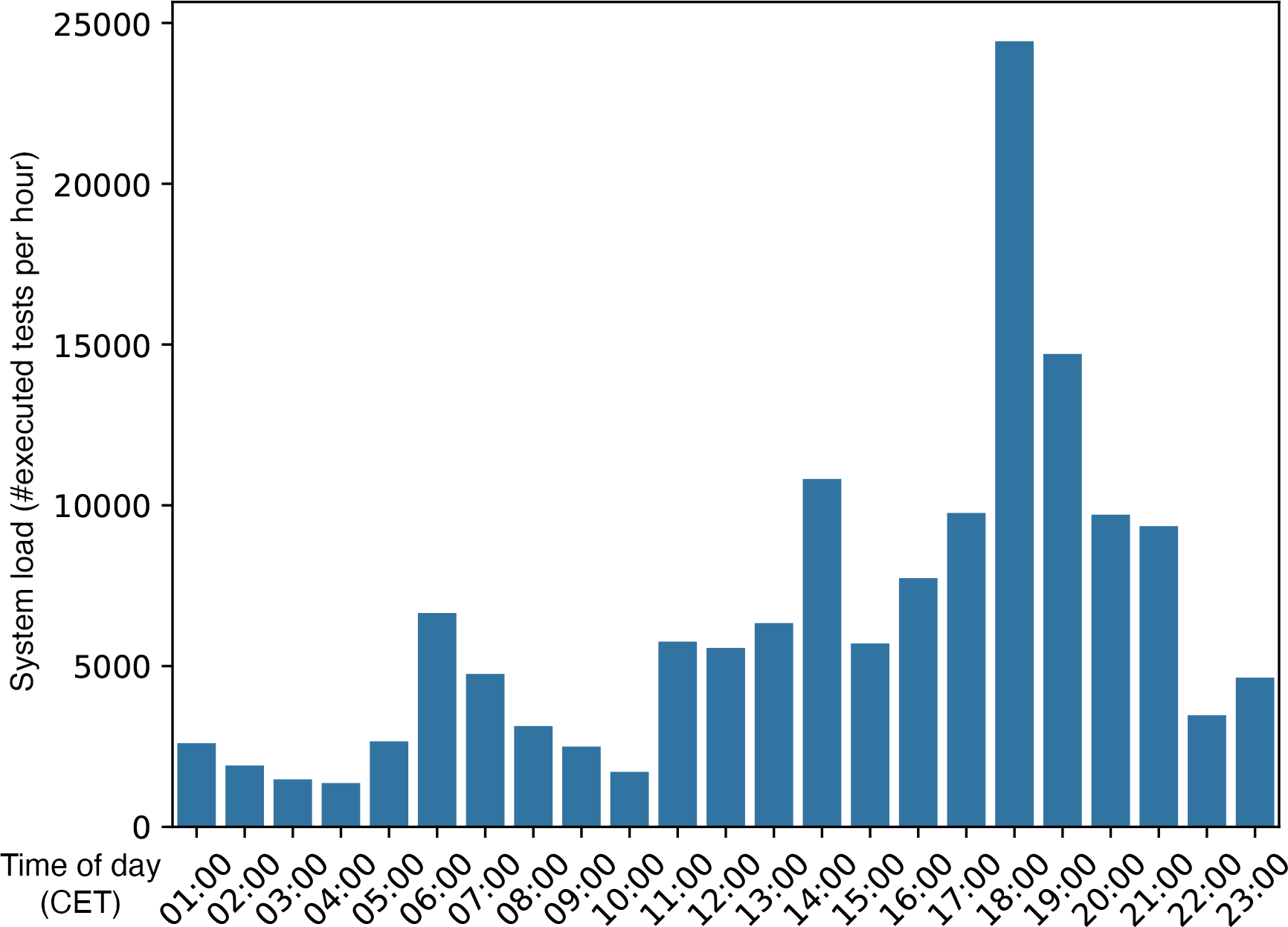}
        \caption{Barplot: Distribution of the system load for a single day.}
        \label{fig:load-example}
    \end{subfigure}
    \caption{Relationship of system load and mean flakiness rate together with system load over an exemplary day.}
    \label{fig:system-load}
    \Description{Scatterplot showing the relationship of system load and mean flakiness rate together with a barplot showing the system load over an exemplary day.}
\end{figure*}

\Cref{fig:load-example} illustrates the idea of this operationalization with the example of a single day in the given dataset. As \Cref{fig:load-example} shows, the peak of the system load on this day occurs at 18:00 (6pm) Central European Time (CET), which is intuitive, as developers typically merge their latest changes before the end of their work day. The number of test executions per hour on that day ranges from \num{1363} at 4:00 (4am) to \num{24443} at 18:00 (6pm) with a mean value of \num{6381}. As SAP's headquarters are located in Germany, most developers work in the CET timezone. However, part of the development is also done in different time zones, which explains the local maximum in the early morning hours. 

To analyze the relationship between system load and flakiness rate, we group the given samples into 100 equal-width bins based on their load, similar to our approach in \Cref{sec:rq1}. We then calculate the average flakiness rate per bin. The resulting average flakiness rates are shown in \Cref{fig:load}.
As \Cref{fig:load} shows, there is no (linear) relationship between the system load and the flakiness rate. In fact, \emph{Pearson's correlation coefficient} is $r_{PE} = 0.08$ with a $p$-value of $0.91$.
Therefore, we conclude that there is no linear relationship between system load and flakiness rate. 

\framed{Answer RQ2 (system load)}{In the data we analyzed, there is no significant correlation between the system load (executed tests per hour) and the flakiness rate.}

\section{Discussion}
\label{sec:discussion}
We discuss the practical implications of our empirical results. 

\textbf{On test execution time}: Our analysis in the context of SAP HANA revealed a significant positive correlation between the average execution time of a test and its flakiness rate. This finding is in line with previous research that identified test execution time as an effective feature to predict whether a test is flaky~\cite{alshammari2021flakeflagger, lampel2021life}. 
%This finding seems intuitive.
As previous research has argued, all tests can be considered flaky~\cite{harman2018start, SAP} because every test might fail flakily due to interruptions of its execution context by some external event. Assuming that such interrupting events are uniformly distributed over time, this would explain a linear relationship between the execution time of a test and its flakiness rate. For example, for tests requiring a network connection, the rate of flaky failures due to network outages grows with the time that they depend on the network during execution. As the test execution time increases, the probability of a network outage occurring at some point during that execution time also increases.

To mitigate the problem caused by this relationship, one common strategy is to split long-running tests to decrease their flakiness~\cite{bach2018effects}. While splitting a single long-running test tends to decrease the flakiness rate, it also increases the effectiveness of re-executing flaky tests, as the scope of re-executions is narrowed down to the actual point of failure. However, executing multiple tests after splitting one larger test might cause additional overhead due to test setup times. Therefore, developers must find an appropriate trade-off between a narrow scope for efficient re-executions of flaky failures and a reduction of the required setup times for the separated tests. 

\textbf{On required computational resources}: Unexpectedly, the relationship between the computational resources required for a test and its flakiness rate varies between the datasets we used in our study. Although a significant positive correlation exists in the PE dataset between the flakiness rate and the required CPU threads or the required memory, respectively, we did not observe significant correlations in the MT dataset. 
We assume that the correlation in the PE dataset is a spurious correlation, which was confounded by a different aspect that we did not consider in this study. Future work is required to investigate potential confounding factors. 

\textbf{On testing distributed environments}: Contrary to our expectations, we found that distributed tests actually showed a lower flakiness rate than non-distributed tests. One possible explanation could be that developers implementing distributed tests have a special focus on the consistency of their tests due to the inherent consistency discussion around distributed systems. % as described in \Cref{sec:questions}.
In fact, the distributed test that yielded the highest flakiness rate in our experiment was fixed during our study period. 

Fixing flaky tests is often a tedious task for developers, as flaky failures can be difficult to reproduce~\cite{lam2020understanding, lam2020large}. As noted in previous research, fixing flaky tests usually has a lower priority compared to tests with permanent failures~\cite{eck2019understanding}. However, due to the consistency requirements of functionality in distributed systems, we expect that resolving or reducing flaky failures in distributed tests will gain increasing attention. 

\textbf{On execution host performance}: In the data we analyzed, we observe a significant positive correlation between our host performance indicator and the flakiness rate. Based on our host performance definition, we conclude that flakiness rates increase when tests are executed on hosts with a lower performance. 
This finding is in line with our finding for \textbf{RQ1}.
The longer tests are executed, the higher their flakiness rate, regardless of the reasons for the long execution time.
Long execution times can be caused by the scope of the test itself or a slow execution environment. 

To mitigate the problems caused by the high variance in execution times on heterogeneous hosts in SAP HANA's testing environment, our previous research on reducing timeout flakiness~\cite{SAP} has motivated the introduction of a global static timeout value of 2 hours for each test execution.
When collecting the data for this study, this global static timeout value was already active. Therefore, we assume that most flaky failures in our study were not caused by flaky timeouts but by other forms of flakiness. % are caused by \emph{async wait} problems. 

\textbf{On system load}: In our analysis, we did not find a significant relationship between the flakiness rate and the load on the testing system. We shared this finding with practitioners at SAP and received the feedback that this might indicate that the load-balancing of SAP HANA successfully distributes tests to prevent exceeding resource limits.

\section{Threats to Validity}
\label{sec:threats}
We discuss threats to the construct, internal, and external validity.

\subsection{Construct Validity}
We discuss the degree to which our operationalizations of the investigated characteristics measure the intended properties~\cite{ralph2018construct}.

\textbf{On required computational resource labels}: As mentioned in \Cref{sec:questions}, SAP HANA developers need to provide a label that specifies the resources required to run a test in terms of available CPU threads and main memory. However, from an organizational perspective, SAP HANA's testing budget limits the runtime (i.e., clock time) that developers can spend running their tests rather than limiting the amount of computational resources the tests consume~\cite{bach2018effects}. As a result, labeling tests with higher resource requirements than needed does not affect the testing budget that developers have. On the contrary: running tests on pools with higher performance might even reduce the test runtime and hence the impact on the testing budget.  
This, in turn, might lead to ``over-sized'' resource labels, which do not reflect the actual required amount of computational resources that the tests need.

\textbf{On the operationalization of system load}: In this study, we approximate the load on SAP HANA's testing infrastructure by counting the number of tests that were executed in a given period, i.e. one hour. This measure does not account for the scope and size of the executed tests, which might bias the results.
However, since the executed test suites are relatively stable over time, we expect the scope and size of the tests to be more or less equally distributed across the intervals.
%Since most test executions are caused by test runs of identical test suites, however, we expect the distribution of test types to be equal across intervals.
Beyond the context of our study, the number of tests executed in a certain period is commonly used as a proxy to monitor the system load at SAP HANA.
However, the chosen interval of one hour could lead to a loss of information since peaks that occur for a short time might not be visible. 

\subsection{Internal Validity}
We discuss the degree to which we can dismiss alternative explanations for our results~\cite{brewer2000research}.

As we draw our conclusions based on an isolated analysis of the flakiness rate and certain characteristics of our tests, our analysis might be vulnerable to confounding variables that influence the examined relationships. To mitigate this threat, we repeated our analyses on two different datasets, one based on production data (PE) and one based on targeted experiments (MT). However, for future work, we encourage studies to investigate more sophisticated relationships and interactions between different test and test environment characteristics.

\subsection{External Validity}
We discuss how our results generalize to other projects~\cite{baltes2022sampling}. 

Due to the specific context of SAP HANA, our results are restricted to this context and may not generalize to other projects. Previous research has also shown that the contributing factors for flakiness vary between different types of projects. Therefore, we encourage further studies targeting other industrial software systems to validate our findings. 

\section{Conclusion}
\label{sec:conclusion}

We conducted a study of test flakiness in the context of a large industrial database management system, SAP HANA. Motivated by previous research and internal discussions at SAP, we tested a range of test and environmental characteristics regarding their relationship with test flakiness. Instead of relying on the traditional binary notion of flakiness, we investigate whether these characteristics increase the flakiness rate of tests. 

Our study reveals that there exists a significant positive correlation between the flakiness rate and the average execution time of a test ($r_{PE}=0.74$, $r_{MT}=0.79$). According to our findings, this is true not only when the scope of a test causes longer execution times but also when it is caused by a slow infrastructure.
Moreover, we found that tests that require distributed setups are significantly less flaky than tests that were executed on a single host. We attribute this finding to the increased priority that distributed systems developers assign to problems related to consistency.

In contrast to a previous industry report by Google, we found the relationship between the flakiness rate of a test and its required resources to be inconclusive. We expect that there exist confounding factors that influence this relationship, and thus encourage future studies to further examine the relationship between flaky tests and the computational resources a test requires. 
Interestingly, we did not find a significant correlation between the load on SAP HANA's testing infrastructure and test flakiness. When we shared this finding with practitioners at SAP HANA, they interpreted it as a confirmation that the load balancer successfully distributes tests across the infrastructure.
Furthermore, this finding leads to the conclusion that the current scale of the testing infrastructure is capable of handling the immense computational demands of SAP HANA's continuous integration pipelines.

Since the results of our study are in the unique context of SAP HANA, we encourage future studies to investigate how our findings generalize to other industrial or open-source software systems. 
%It might also be interesting to relate test flakiness to existing software quality standards for identifying software quality characteristics that correlate with flakiness. 

\bibliographystyle{ACM-Reference-Format}
\bibliography{literature}
\end{document}